\documentclass[conference]{IEEEtran}
\IEEEoverridecommandlockouts
\usepackage{cite}
\usepackage{amsmath,amssymb,amsfonts}
\usepackage{algorithmic}
\usepackage{graphicx}
\usepackage{textcomp}
\usepackage{xcolor}
\def\BibTeX{{\rm B\kern-.05em{\sc i\kern-.025em b}\kern-.08em
    T\kern-.1667em\lower.7ex\hbox{E}\kern-.125emX}}
\begin{document}

\title{A Compact 3D-Printed Soft Finger with\\ Cyclic Hydraulic Actuation}

\author{
Zefang Mao$^{1}$, Yunjia Li$^{1}$, Victoria Bamgboye$^{1}$, Sara Ben Haj Hammouda$^{1}$, Aika Ono$^{1}$, Wen Fan$^{1}$,\\ Chao Wu$^{2}$ and Dandan Zhang$^{1,*}$%
\thanks{
$^{1}$Department of Bioengineering, Imperial College London, London, United Kingdom.
$^{2}$Department of Civil and Environmental Engineering, Imperial College London, London, United Kingdom.
$^{*}$Corresponding author: Dandan Zhang, \texttt{d.zhang17@imperial.ac.uk}.
}%
}

\maketitle

\begin{abstract}
Hydraulic soft fingers offer compliant and gentle manipulation, but their practical deployment is limited by bulky fluidic hardware, fabrication complexity, and insufficient design validation. This paper presents a compact 3D-printed soft hydraulic finger driven by a miniature cyclic peristaltic loop. The finger integrates compliant bellows, rigid connectors, and fluidic ports, while an Abaqus fluid-structure model is used to guide selection of wall thickness, pitch angle, and bellows length. The selected design is validated through baseline-corrected chamber-pressure measurements and vision-based angle tracking. Results show that the quasi-static finite-element model captures the main pressure-angle trends, with remaining offsets mainly attributed to bonding-induced stiffness and hydraulic losses. Vision-feedback control further enables repeatable angle tracking over a large bending range. Finally, grasping tests on fragile and deformable objects, including tofu and blueberries, demonstrate gentle, slip-free contact without visible damage. Overall, this work establishes a reproducible pipeline from FEA-guided design to closed-loop validation for compact 3D-printed hydraulic soft fingers.
\end{abstract}


\section{Introduction}

Soft robotic grippers are well suited to manipulating irregular, fragile, and deformable objects, as their compliance enables passive shape adaptation and distributed contact forces~\cite{hughes2016soft}. Among available strategies, fluidic actuation~\cite{shi2022review} is widely used to generate large, reversible deformation. Pneumatic systems are lightweight and safe, but air compressibility can introduce delays and pressure fluctuations~\cite{xavier2021design}. By contrast, hydraulic actuation uses a nearly incompressible fluid, enabling stable pressure transmission, higher force density, and gentle contact with fragile objects.

Despite these advantages, hydraulic soft fingers face two practical challenges. First, fabrication and sealing require enclosed chambers, robust walls, and reliable fluidic interfaces that withstand repeated pressurisation without leakage or failure. Conventional moulding and casting remain common, but are labour-intensive, alignment-sensitive, and slow to iterate~\cite{wang2020dual,galloway2013mechanically}, especially for small chambers, embedded connectors, and complex bellows geometries. Second, traditional hydraulic systems often rely on external pumps, reservoirs, valves, and long fluidic lines, increasing footprint, dead volume, latency, and routing complexity~\cite{shi2022review,katzschmann2016cyclic}. These limitations hinder integration into compact grippers.

Additive manufacturing offers a route to compact soft structures with integrated fluidic features. In particular, PolyJet printing~\cite{vijayan2021evaluation,yap2020review,xin2024role} can co-print soft and rigid photopolymers at relatively high resolution, integrating compliant chambers and rigid connectors within a single design. However, miniaturised hydraulic fingers still require geometric optimisation: thin walls improve bending compliance but increase stress concentration and leakage risk, whereas thicker structures improve robustness at the expense of bending efficiency. Finite element analysis (FEA) is therefore essential for relating geometry to deformation, stress distribution, and actuator performance before fabrication~\cite{demir2020computational}.

A compact hydraulic architecture is also required. Although dual-chamber soft fingers can provide more controllable bending than single-chamber designs, independent chamber control typically requires multiple pumps, valves, tubes, and interfaces. Cyclic hydraulic actuation~\cite{maccurdy2016printable,katzschmann2016cyclic} instead uses closed-loop fluid exchange between chambers, enabling bidirectional bending with a single recirculating loop and reducing hardware burden, routing complexity, and footprint. Nevertheless, pressure-deformation behaviour must be experimentally validated, and open-loop pump commands remain insufficient because residual bubbles, hydraulic losses, and soft-structure dynamics introduce command-to-angle variability. A compact platform combining pressure measurement, bending-angle tracking, and closed-loop control is therefore required.








Recent work has explored two routes towards compact soft grippers. Bell et al.~\cite{bell2022modular} integrated the pump, flow lines, and actuator into a self-contained closed hydraulic system driven by a stepper-motor peristaltic pump, reducing external regulators, valves, sensors, and long tubing. However, their soft appendage still relies on mould-based fabrication. In parallel, Meitani and Nisar~\cite{meitani2025materials} demonstrated directly 3D-printed soft grippers using low-hardness TPU; however, their approach uses single-material FDM printing and pneumatic actuation, limiting multi-material integration and differing from cyclic hydraulic operation.

To address these challenges, we present a compact 3D-printed cyclic hydraulic soft finger for gentle grasping. The finger adopts a PolyJet-printed bellows-inspired dual-chamber design, with an Abaqus FEA fluid-structure model guiding the selection of wall thickness, pitch angle, and bellows length. The prototype is characterised through baseline-corrected chamber-pressure measurements, vision-based angle tracking, closed-loop bending control, and grasping tests on fragile objects. The main contributions of this paper are:
\begin{itemize}
\item A compact 3D-printed hydraulic soft finger integrating compliant bellows, rigid connectors, and fluidic ports for cyclic hydraulic actuation.
\item An FEA-guided workflow relating pitch angle, wall thickness, and bellows length to bending efficiency and stress concentration.
\item Vision-feedback closed-loop bending control and fragile-object grasping, using a two-angle descriptor to characterise non-uniform contact-induced bending.
\end{itemize}


\section{Methodology}

\subsection{Soft Finger Design}

\begin{figure}[htbp!]
  \centering
  \includegraphics[width=\columnwidth,keepaspectratio]{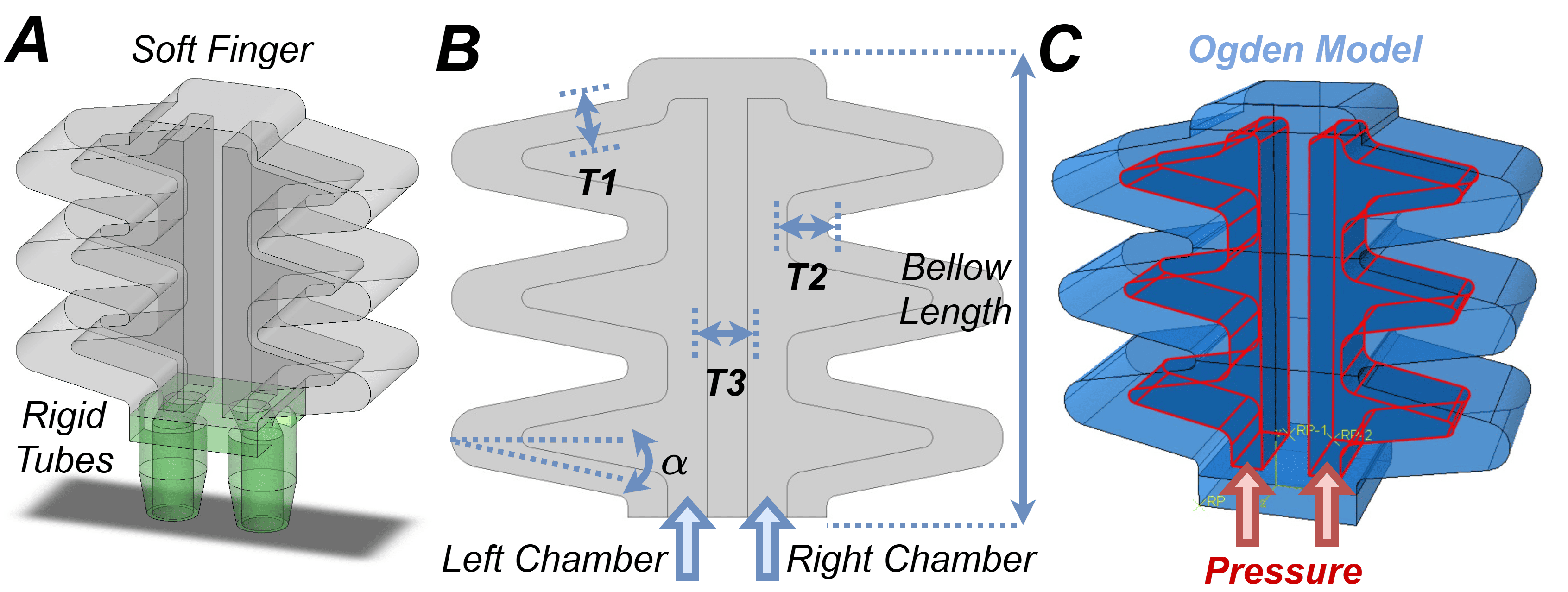}
  \caption{
    Design and FEA setup of the 3D-printed soft hydraulic finger.
    A: 3D model showing the Agilus30 bellows body and Vero tubes/connectors.
    B: Cross-section of the dual-chamber structure with wall thicknesses \(T_1\), \(T_2\), \(T_3\), bellows length, and pitch angle \(\alpha\).
    C: Pressurised FEA model using an Ogden hyperelastic material description.
}
  \label{fig_1}
\end{figure}

As shown in Fig.~\ref{fig_1}(A), the proposed actuator adopts a bellows-inspired dual-chamber structure, with two fluidic chambers separated by a central partition. During cyclic hydraulic actuation, fluid is transferred between the chambers, causing one chamber to expand while the other contracts. The resulting pressure difference between right- and left-chamber pressures, $P_{\mathrm{right}}$ and $P_{\mathrm{left}}$, drives bending towards the lower-pressure chamber, while balanced pressures keep the finger near its neutral straight configuration.

The finger is designed for multi-material PolyJet printing, with Agilus30 used for the compliant bellows body and Vero for the rigid tubes. This material distribution localises deformation in the soft bellows region while providing robust fluidic interfaces, avoiding fibre reinforcement and supporting compact hydraulic routing.

As shown in Fig.~\ref{fig_1}(B), five geometric parameters are considered: wall thicknesses $T_1$, $T_2$, and $T_3$, bellows length, and pitch angle $\alpha$. These parameters determine the trade-off between bending efficiency, structural robustness, and sealing reliability. Thinner side walls increase compliance but may raise local stress concentration, whereas a thicker central partition improves inter-chamber stability.

Finite-element simulations were then used to guide geometric selection. Because soft robotic structures exhibit nonlinear deformation due to large strain, hyperelastic material behaviour, and fluid-structure interaction, the relationship between design parameters and actuator performance cannot be directly derived analytically. Therefore, representative candidate configurations were generated by varying key parameters and compared under consistent loading and boundary conditions. The selected geometry was determined by comparing the bending angle under the same pressure, stress distribution, sealing reliability, and manufacturability.

\begin{figure*}[h!]
    \centering
    \includegraphics[width=0.95\linewidth]{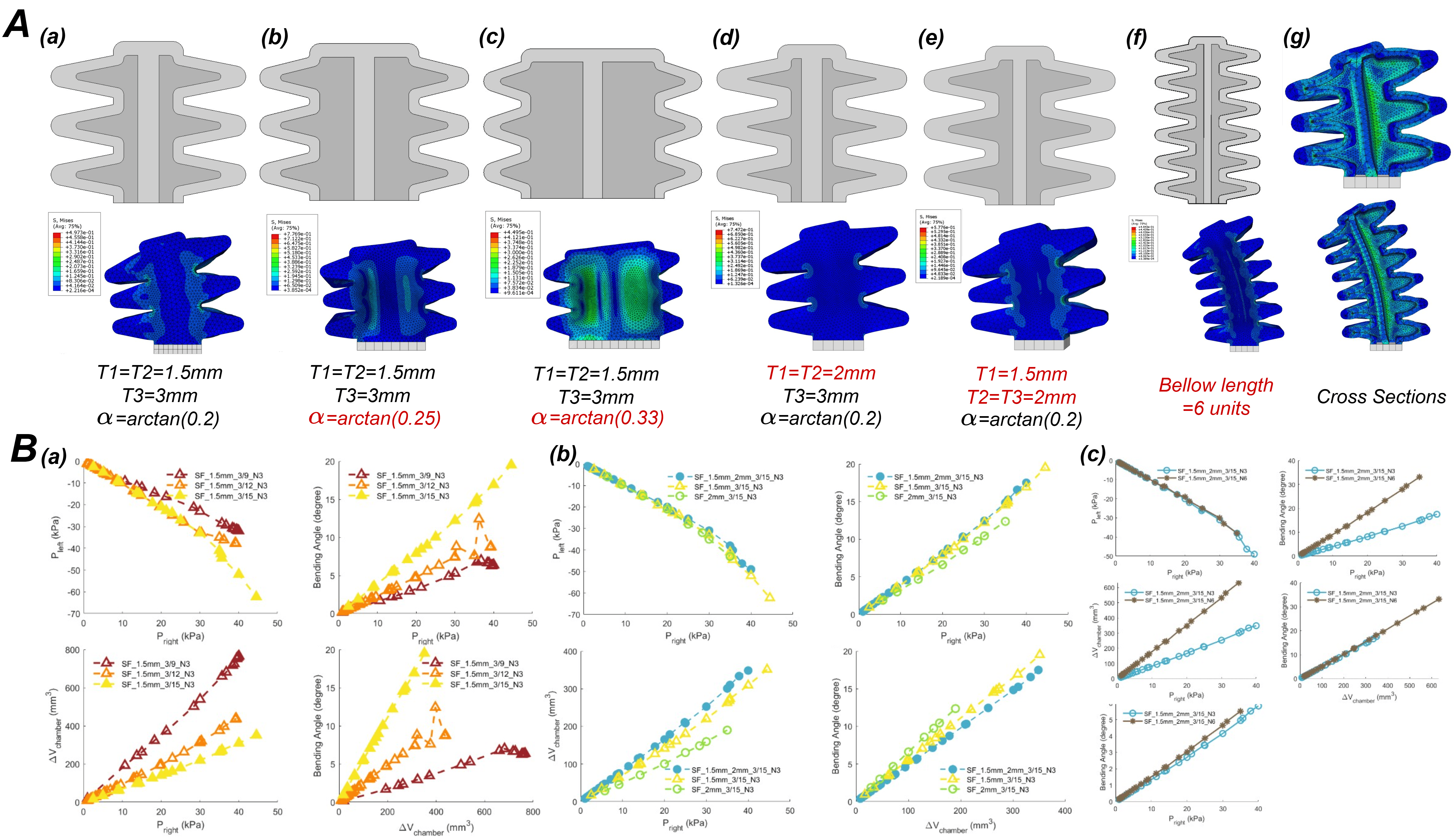}
    \caption{
        FEA-guided comparative screening of representative soft-finger geometries. A: Candidate geometries and simulated stress distributions under \(P_{\mathrm{right}}=35~\mathrm{kPa}\), showing variations in pitch angle, wall-thickness distribution, and bellows length. B: Quantitative comparison of chamber-pressure coupling, pressure-angle response, pressure-volume response, and volume-angle response across the representative configurations.
    }
    \label{fig_2}
\end{figure*}

\begin{table}[h!]
\caption{Ogden hyperelastic material parameters for Agilus30~\cite{abayazid2020material}.}
\centering
\setlength{\tabcolsep}{3pt}          
\renewcommand{\arraystretch}{1.1}      
\resizebox{\columnwidth}{!}{
\begin{tabular}{|c|c|c|c|c|c|c|}
\hline
\textbf{Material} & \multicolumn{6}{|c|}{\textbf{Ogden quasi-static hyperelastic parameters}} \\
\cline{2-7}
 & \textbf{\textit{$\mu_{1}(\infty)$ (MPa)}} & \textbf{\textit{$\alpha_{1}$}} &
\textbf{\textit{$\mu_{2}(\infty)$ (MPa)}} & \textbf{\textit{$\alpha_{2}$}} &
\textbf{\textit{$\mu_{3}(\infty)$ (MPa)}} & \textbf{\textit{$\alpha_{3}$}} \\
\hline
Agilus30 & 0.2127 & 1.3212 & 0.0375 & 4.318 & -0.001 & -1.0248 \\
\hline
\end{tabular}
}
\label{tab:ogden_params}
\end{table}

\subsection{FEA Simulation}

Finite-element analyses (FEA) were conducted in Abaqus to evaluate the deformation, stress distribution, and fluid-structure response of the soft finger. As shown in Fig.~\ref{fig_1}(C), the Agilus30 body was modelled as a nearly incompressible hyperelastic solid using a three-term Ogden model, while the Vero inserts were treated as linear elastic rigid-like components. Near-incompressibility was handled using hybrid elements, with \(J \approx 1\). The strain-energy density is
\begin{equation}
W(\lambda_1,\lambda_2,\lambda_3)
= \sum_{i=1}^{3}\frac{2\mu_i}{\alpha_i^{2}}
\left(
\lambda_1^{\alpha_i}
+\lambda_2^{\alpha_i}
+\lambda_3^{\alpha_i}
-3
\right),
\end{equation}
where \(\lambda_k\) are the principal stretches, and \((\mu_i,\alpha_i)\) are the Ogden parameters calibrated from quasi-static Agilus30 data, as listed in Table~\ref{tab:ogden_params}~\cite{ogden2004fitting,abayazid2020material}. The soft body was discretised using quadratic hybrid tetrahedral elements with a global size of \(\leq 1\,\mathrm{mm}\), while the rigid tubes used linear elastic hexahedral elements. The base was fully fixed, self-contact was frictionless, and the two chambers were modelled as water-filled fluid cavities initialised at \(P=0\,\mathrm{kPa}\).

Because the peristaltic pump imposes coupled fluid exchange, direct pump simulation was approximated using pressure-driven fluid-cavity analyses followed by volume-conservation filtering. The chamber-volume changes are:
\begin{equation}
\Delta V_R=V_R(t)-V_R(0),
\quad
\Delta V_L=V_L(t)-V_L(0).
\end{equation}
A state was accepted when the chambers showed opposite-signed volume changes and less than \(3\%\) mismatch:
\begin{equation}
\mathrm{sign}(\Delta V_R)=-\mathrm{sign}(\Delta V_L),
\end{equation}
\begin{equation}
\frac{
\left|
|\Delta V_R|-|\Delta V_L|
\right|
}{
\min\left(|\Delta V_R|,|\Delta V_L|\right)
}
<0.03 .
\end{equation}
This filtering approximated cyclic, constant-volume hydraulic actuation. Parameter sweeps were performed over pitch angle, wall thickness, and bellows length. The geometric study was not intended to exhaustively optimise a continuous design space. Instead, representative candidates were selected to span the main design trade-offs under the practical constraints. For each accepted state, chamber pressure, bending angle, volume change, and stress distribution were recorded.

\subsubsection{Comparison of Pitch Angle}

The pitch-angle candidates were selected as representative low, medium, and high corrugation levels within a manufacturable range, rather than as an exhaustive search over all possible angles. As shown in Fig.~\ref{fig_2}(A.a--c), the pitch angle was varied using tangent values of 3/15 (0.20), 3/12 (0.25), and 3/9 (0.33), while keeping finger height and wall thickness unchanged. At \(P_{\mathrm{right}}=35\,\mathrm{kPa}\), the 3/9 design produced limited bending and central-body stress concentration. The 3/12 design increased curvature but still showed partially centralised stress. In contrast, the 3/15 design localised deformation within the corrugations, producing larger bending and a more distributed stress field.

The quantitative comparison in Fig.~\ref{fig_2}(B.a) further shows that the 3/15 design achieved the highest pressure-angle gain. Since pressure-volume and volume-angle responses are affected by different initial chamber volumes, they were treated as secondary indicators. Therefore, among the three candidates, the 3/15 configuration was selected for higher bending efficiency, reduced central stress concentration, and stable large-angle deformation.

\subsubsection{Comparison of Wall Thicknesses}

Wall thickness was compared using three representative configurations. The thin-wall design \((T_1=T_2=1.5~\mathrm{mm},~T_3=3.0~\mathrm{mm})\) was used as the high-compliance baseline. The uniformly thickened design \((T_1=T_2=2.0~\mathrm{mm},~T_3=3.0~\mathrm{mm})\) was introduced to assess the effect of increasing the side-wall stiffness. The mixed-thickness design \((T_1=1.5~\mathrm{mm},~T_2=T_3=2.0~\mathrm{mm})\) was then used to reduce stress concentration and improve sealing while maintaining side-wall compliance. In the mixed design, \(T_3\) was reduced from 3.0 mm to 2.0 mm to avoid excessive chamber narrowing. 

From Fig.~\ref{fig_2}(B.b), the thin-wall design produced large deformation at \(P_{\mathrm{right}}=35\,\mathrm{kPa}\) but concentrated stress at the corrugation valleys and central web. Uniform thickening reduced stress concentration but increased stiffness and suppressed curvature. The mixed-thickness design redistributed stress while maintaining bending angles comparable to the thin baseline, and was selected as a compromise between bending efficiency, stress reduction, and sealing reliability.

\subsubsection{Comparison of Bellows Length}

As shown in Fig.~\ref{fig_2}(A.f,g), three-unit and six-unit designs were compared with fixed pitch angle and wall-thickness distribution to examine length scaling of the selected unit-cell geometry. The shorter design showed more uneven stress distribution, whereas the six-unit design produced a more uniform internal stress field. The pressure-angle curves in Fig.~\ref{fig_2}(B.c) show that the six-unit design reached larger bending angles under the same pressure because of greater compliance and chamber volume. After length normalisation, both designs showed comparable bending efficiency, while the six-unit design provided a larger deformation range.

\begin{figure}[h!]
    \centering
    \includegraphics[width=0.95\linewidth]{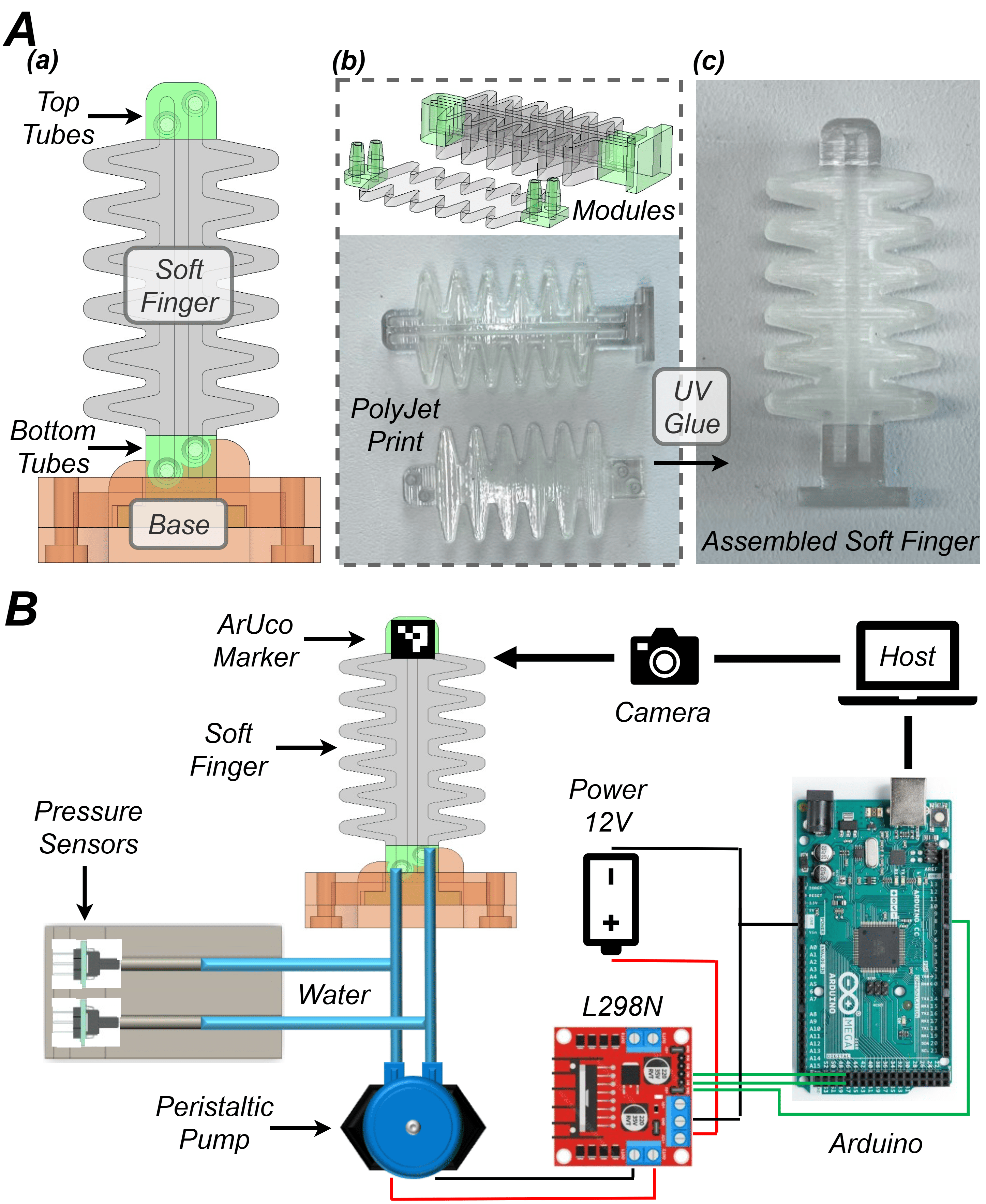} 
    \caption{
        Fabrication and experimental setup of the soft hydraulic finger.
        A: Final design with top/bottom tubes, rigid base, and split PolyJet modules bonded using UV-curable resin.
        B: Cyclic hydraulic testbed with peristaltic pump, pressure sensors, Arduino-L298N control, and ArUco-based tracking.
    } 
    \label{fig_3}
\end{figure}

\subsection{PolyJet Printing}

Based on the comparative FEA screening,   the unit-cell geometry \(T_1=1.5\,\mathrm{mm},\,T_2=T_3=2.0\,\mathrm{mm}, tan(\alpha)=3/15\) was selected to balance bending efficiency, stress concentration, and footprint. Paired top and bottom ports were integrated to improve hydraulic reliability: the bottom ports connected to the hydraulic loop, while the top ports enabled filling and bubble removal before being sealed with short connectors and plugs. A rigid base was added for secure fixation during bench-top testing, as shown in Fig.~\ref{fig_3}(A.a).

Then, the actuator was fabricated by multi-material PolyJet printing. Because support material had to be removed from the enclosed corrugated chambers, a fully monolithic print was impractical. The finger body was therefore printed as split modules along the horizontal mid-plane and bonded using UV-curable resin (Fig.~\ref{fig_3}(A.b/c)). This strategy preserved the dual-chamber geometry and enabled support removal, but introduced a bonded mid-plane that was expected to increase stiffness and affect sealing.

\subsection{Fluidic Actuation and Sensing}

Fig.~\ref{fig_3}(B) shows the closed hydraulic testbed used to validate the bending response of the 3D-printed finger. Water was used as the working fluid because of its safety and near-incompressibility, enabling comparison with the quasi-static FEA model. The two chambers were connected to a bidirectional recirculating loop driven by a miniature peristaltic pump (NKP-DC-S10B, 12~V, 5.2-90~mL/min). Reversing the pump direction swaps the pressurised and depressurised chambers, enabling left and right bending without additional valves.

To keep the system compact, inline hydraulic transducers were avoided. Instead, each fluidic branch included a short horizontal column with a trapped air pocket connected to a pneumatic pressure sensor. Since the trapped air remains in static equilibrium with the liquid, the measured gauge pressure estimates the corresponding chamber pressure. This arrangement maintains a closed hydraulic circuit while providing compact pressure readout during cyclic actuation.

Finger bending was measured independently using vision feedback. An ArUco marker was attached to the fingertip and tracked by a monocular camera to estimate the global bending angle. Pump speed and direction were commanded by an Arduino through an L298N H-bridge driver, while pressure data and camera frames were time-stamped and logged on a laptop. During operation, bending is generated by simultaneously pressurising one chamber and depressurising the other. The horizontal fluid columns visualise this pressure exchange: the air pocket is compressed on the pressurised side and expands on the depressurised side. All chamber pressures are reported as baseline-corrected values,
\(\Delta P(t)=P(t)-P(0)\), where \(P(0)\) is the initial neutral reading before actuation.

\section{Experiment}

\subsection{FEA-to-Experiment Validation}

\begin{figure}[h!]
    \centering
    \includegraphics[width=0.95\linewidth]{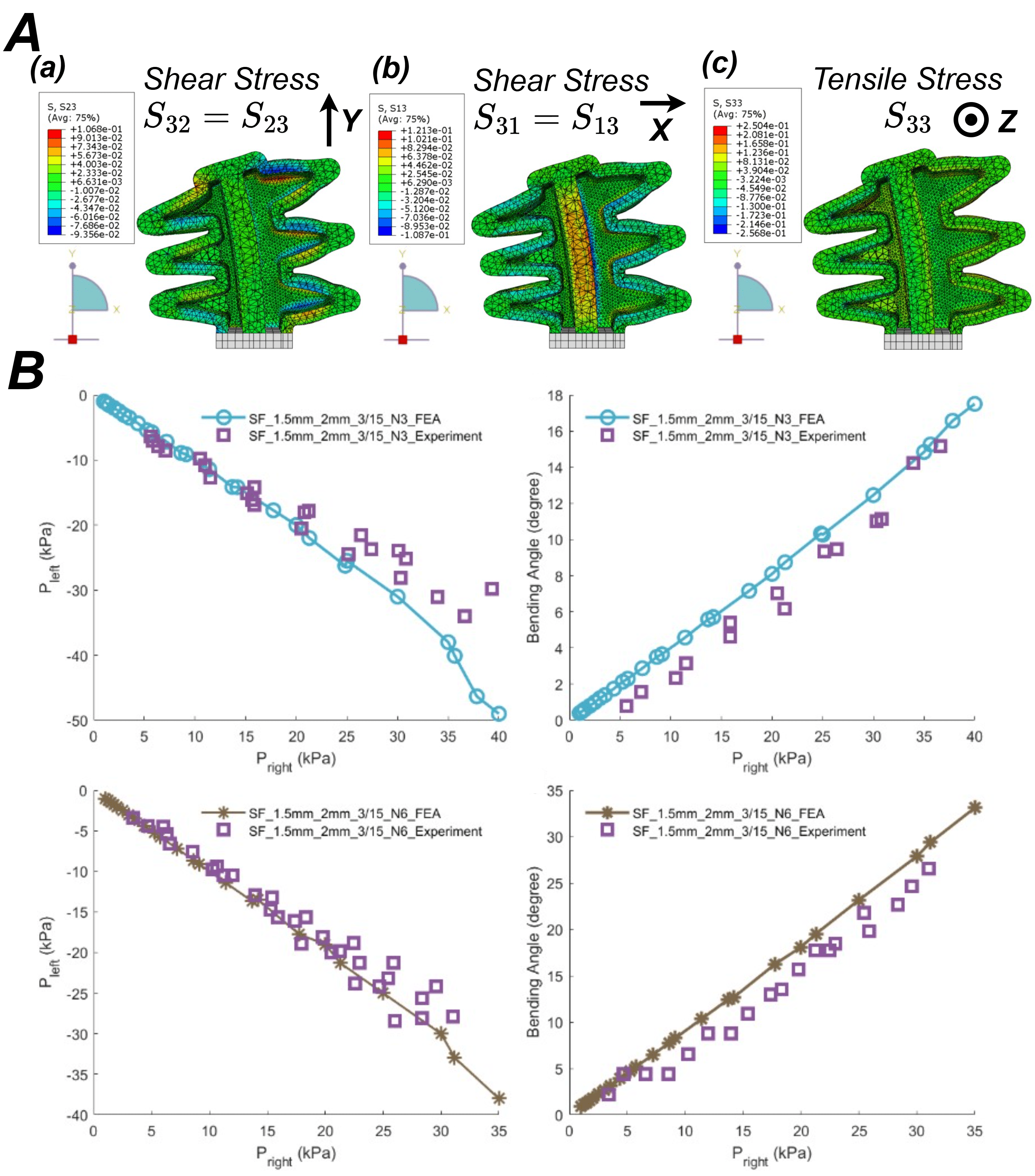}
    \caption{
    FEA stress analysis and experimental validation.
    A: Simulated stress components showing concentrations along corrugated walls and the bonded mid-plane.
    B: FEA-experiment comparison of chamber-pressure coupling and pressure-angle responses for three- and six-unit fingers.
    }
    \label{fig_4}
\end{figure}

The simulated stress fields were analysed to identify potential failure modes in the printed actuator. As shown in Fig.~\ref{fig_4}(A), shear stress \(S_{23}\) concentrates along the corrugated fold walls, indicating possible shear-driven sidewall damage. Shear stress \(S_{13}\) is localised near the bonded mid-plane and central partition, suggesting a potential inter-chamber debonding path. Tensile stress \(S_{33}\) appears across the adhesive seam, indicating possible delamination under repeated pressurisation. These results identify inter-chamber leakage and seam delamination as the main risks of the split-and-bond strategy.

Fig.~\ref{fig_4}(B) compares FEA predictions with experimental measurements for the three-unit and six-unit fingers. For both designs, the experiments follow the FEA trends: the two chambers exhibit an inverse pressure relationship during cyclic hydraulic actuation, and the bending angle increases linearly with \(\Delta P_{\mathrm{right}}\).

Systematic offsets are still observed. At higher loads, the measured negative pressure in the opposing chamber is smaller than predicted, and the experimental bending angles are consistently lower than the numerical results. This is mainly attributed to the bonded mid-plane, which increases stiffness and constrains deformation relative to the idealised FEA model. Minor viscous losses and residual air bubbles may further contribute to the discrepancy.

Despite these offsets, the agreement in overall trend indicates that the quasi-static FEA captures the dominant chamber-coupling and bending behaviour, providing a useful basis for predicting pressure--deformation responses and guiding geometry optimisation.

\begin{figure*}[h!]
    \centering
    \includegraphics[width=0.95\linewidth]{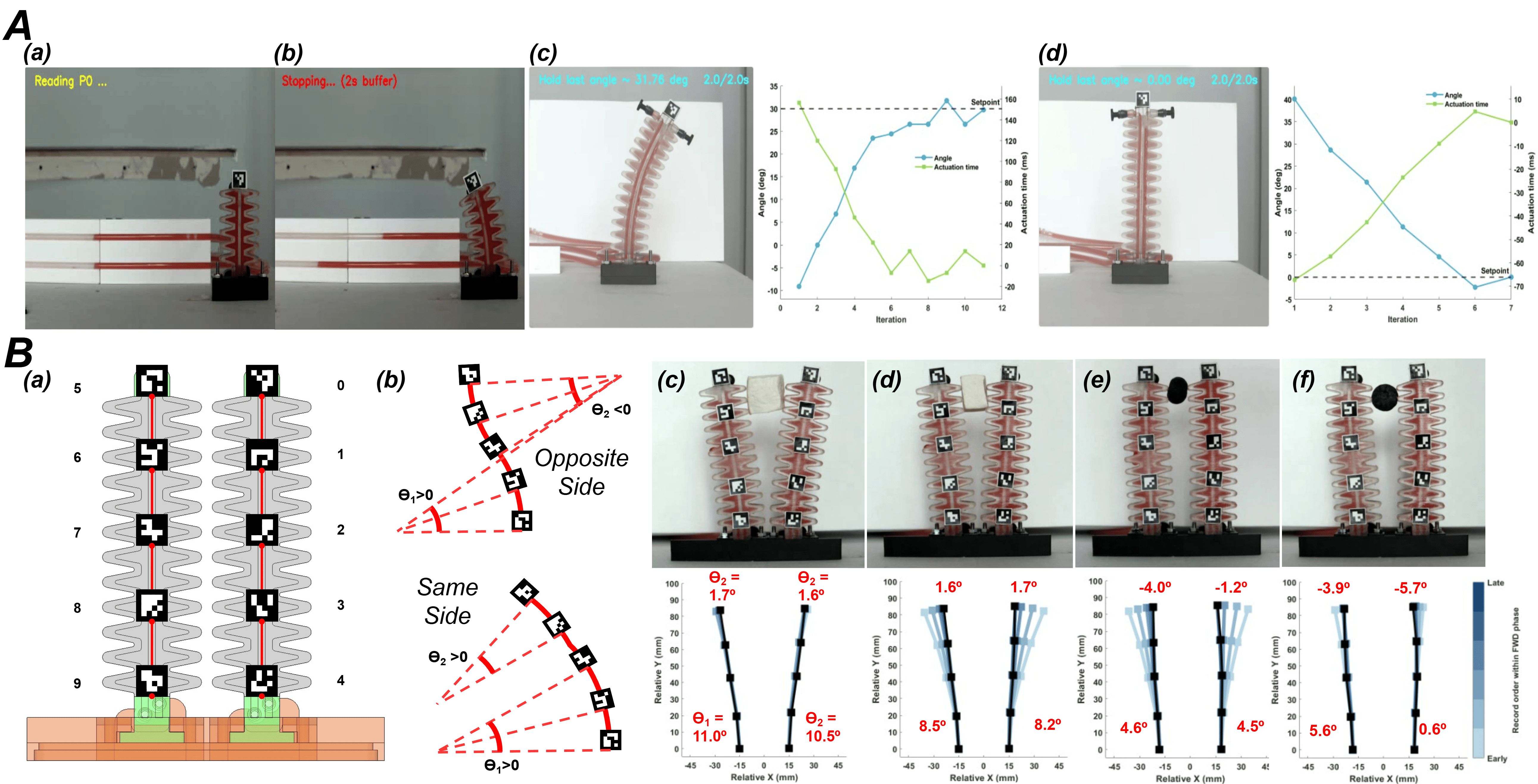}
    \caption{
    Closed-loop control and grasping characterisation.
    A: Vision-feedback bending control with snapshots and angle/actuation-time tracking.
    B: Multi-marker centreline tracking for extracting \(\theta_1\) and \(\theta_2\) during contact with tofu cubes and blueberries.
    }
    \label{fig_5}
\end{figure*}

\subsection{Closed-Loop Tracking}

For FEA-to-experiment validation, three-unit and six-unit prototypes were used to evaluate length scaling of the selected unit-cell geometry while keeping simulation and fabrication complexity manageable. After fixing the pitch angle and wall-thickness distribution, the same unit-cell design was extended to a 12-unit finger for closed-loop tracking and grasping tests. This longer finger was not treated as a separately optimised geometry, but as an application-scale extension of the validated six-unit design to provide a larger bending range and grasping workspace. However, the longer configuration also amplified mechanical compliance, fluidic delays, and command-to-angle variability caused by residual bubbles, hydraulic losses, sensing latency, and soft-structure settling. Vision-feedback control was therefore implemented to regulate the final bending angle.

Fig.~\ref{fig_5}(A.a,b) illustrates the visual feedback from the coloured fluid columns. During actuation, fluid transfer between chambers causes the red columns to move in opposite directions: the pressurised side extends as the trapped air pocket is compressed, while the depressurised side shortens as the air pocket expands. This liquid-level change verifies cyclic hydraulic actuation, but is not precise enough for soft finger control.

From Fig.~\ref{fig_5}(A.c,d), the bending angle was estimated from the camera-tracked ArUco marker, and the target error was converted into a bounded pump actuation duration. The error sign determined the flow direction: positive commands pressurised the right chamber, whereas negative commands pressurised the left chamber. Once the measured angle entered a \(\pm 1^\circ\) tolerance band, the pump was stopped and the finger was held for \(2\,\mathrm{s}\) to record the stabilised final state. Each command was followed by a \(0.5\,\mathrm{s}\) settling interval for pressure and angle logging.

The controller was evaluated on a 12-unit finger using two target motions. For a step command to \(30^\circ\), the angle reached the target within 11 iterations (Fig.~\ref{fig_5}(A.c)), while returning from a bent configuration to \(0^\circ\) required 7 iterations (Fig.~\ref{fig_5}(A.d)). Small fluctuations near the target were attributed to structural compliance and pump timing resolution, but the final angle remained within the \(\pm 1^\circ\) tolerance band.

\subsection{Grasping Characterisation}

For grasping tests, two identical 12-unit soft fingers, obtained by extending the validated unit-cell geometry, were mounted symmetrically on a rigid base to form a two-finger gripper. Each trial followed the same cyclic actuation protocol: a short pump pulse followed by a \(0.5\,\mathrm{s}\) measurement pause. From the neutral configuration, the fingers were first actuated outward for \(N_{\mathrm{out}}\) iterations to create an open pre-shape, and then reversed for inward bending. At the start of the inward phase, a tofu cube or blueberry was placed near the upper central region of the gripper and released once first contact occurred.

To describe non-uniform bending beyond single-tip-angle measurement, a two-angle representation was introduced. Multiple ArUco markers were placed along the finger (Fig.~\ref{fig_5}(B.a)), and the midpoints of their lower edges defined a centreline. As shown in Fig.~\ref{fig_5}(B.b), two local bending angles were extracted: \(\theta_1\) for the lower segment and \(\theta_2\) for the upper segment. Same-signed angles indicate single-arc in-plane bending, whereas opposite signs indicate an inflexion and curvature change.

\subsubsection{\textit{Flat objects}} 
When grasping tofu cubes (Fig.~\ref{fig_5}(B.c,d)), \(\theta_1\) and \(\theta_2\) generally had the same sign, with \(|\theta_1| \gg |\theta_2|\). This indicates planar contact dominated by lower-segment deformation. Side grasping reduced the apparent object width, leading to a smaller \(\theta_1\) with little change in \(\theta_2\).

\subsubsection{\textit{Spherical objects}} 
For blueberries (Fig.~\ref{fig_5}(B.e,f)), \(\theta_1\) and \(\theta_2\) often had opposite signs, indicating an inflexion that conformed to the curved surface. Side contact produced a smaller \(|\theta_2|\), whereas frontal contact increased both \(|\theta_2|\) and \(\theta_1\) to accommodate the larger projected width.

Overall, planar contacts were characterised by large \(\theta_1\) and near-zero \(\theta_2\), whereas curved contacts induced opposite-signed angles and an inflexion. Across the tested cases, tofu and blueberries were grasped without visible damage or slip. The two-angle descriptor \((\theta_1,\theta_2)\) therefore provides a compact representation of contact-induced non-uniform bending for gentle grasping.

\section{Conclusion}

This paper presented a compact PolyJet-printed soft finger driven by cyclic hydraulic actuation. An FEA-guided workflow was used to compare representative pitch angles, wall-thickness distributions, and bellows lengths by balancing bending efficiency, stress concentration, sealing reliability, and manufacturability. The selected unit-cell geometry was fabricated from split PolyJet-printed modules and validated through baseline-corrected pressure sensing and vision-based angle tracking. Results showed that the quasi-static FEA captured the main chamber-coupling and pressure-angle trends, with residual offsets mainly attributed to bonding-induced stiffness, hydraulic losses, and air bubbles. After extending the validated unit-cell geometry to a 12-unit application-scale finger, vision-feedback control enabled repeatable tracking over a large motion range. Grasping experiments with tofu cubes and blueberries showed that the finger adapted to flat and curved fragile objects without visible damage or slip. The two-angle descriptor \((\theta_1,\theta_2)\) compactly represented contact-induced non-uniform bending beyond single-tip-angle measurement. The main limitations arise from the bonded mid-plane, residual bubbles, and the settling interval between actuation steps. Future work will expand the geometric design space and evaluate multi-finger configurations.

\vspace{12pt}

\bibliographystyle{IEEEtran}  
\bibliography{reference}

\end{document}